\documentclass[12pt,preprint]{aastex}
\usepackage{lscape}

\newcommand{\bdv}[1]{\mbox{\boldmath$#1$}}

\def\rel{{\rm rel}}
\def\e{{\rm E}}
\def\au{{\rm AU}}

\def\kms{{\rm km}\,{\rm s}^{-1}}

\def\rel{{\rm rel}}
\def\bv{{\bf v}}

\def\bs{{\bf s}}
\def\bn{{\bf \hat n}}
\def\be{{\bf \hat e}}
\def\e{{\rm E}}
\def\bpi{{\bdv{\pi}}}
\def\balpha{{\bdv{\alpha}}}
\def\bmu{{\bdv{\mu}}}
\def\bomega{{\bdv{\omega}}}
\def\bLambda{{\bdv{\Lambda}}}
\def\bepsilon{{\bdv{\epsilon}}}
\def\bu{{\bf u}}
\def\bj{{\bf j}}
\def\ec{{\rm ec}}

\begin{document}
\title{Resolution of the MACHO-LMC-5 Puzzle: The Jerk-Parallax 
Microlens Degeneracy}

\author{Andrew Gould}

\affil{
Department of Astronomy, The Ohio State University, 
140 West 18th Avenue, Columbus, OH 43210\\
gould@astronomy.ohio-state.edu
}

%\submitted{Submitted to The Astrophysical Journal}

\begin{abstract}

By extending the constant-acceleration analysis of Smith, Mao, \& Paczy\'nski
to include jerk, I show that microlens parallax measurements are subject
to a four-fold discrete degeneracy.  The new degeneracy is characterized
by a projected velocity 
$\tilde v_j = -(3/4)\csc\beta_\ec(\cos^2\psi\sin^2\beta_\ec+
\sin^2\psi)^{3/2}v_\oplus$,
where $\beta_\ec$ is the ecliptic latitude, $\psi$ is the phase of the Earth's
orbit relative to opposition at the time of the event maximum, and 
$v_\oplus=30\,\kms$ is the speed of the Earth.  The degeneracy becomes
important when the lens projected velocity $\tilde v$ is of order
$\tilde v_j$.  For events toward the Large Magellanic Cloud, 
$\tilde v_j\simeq (3/4)v_\oplus$,
so this degeneracy is important primarily for lenses in the Milky Way
disk.  In particular, it solves the puzzle of MACHO-LMC-5, 
whose microlens parallax measurement had yielded mass and distance
determinations for the lens that were inconsistent with photometric
estimates.  Toward the Galactic bulge, $\tilde v_j$ ranges from 
$\sim 0.2\,\kms$ at the solstice to $\sim 200\,\kms$ at the
equinoxes, so the effect of the degeneracy depends strongly on the peak
time of the event.  The degeneracy applies mainly to events with
Einstein timescales, $t_\e\la {\rm yr}/2\pi$.

\end{abstract}


\section{Introduction
\label{intro}}

The spectacular high-magnification 
microlensing event MACHO-LMC-5 has been a puzzle
since its discovery was reported by the MACHO collaboration \citep{alcock97}.
The ``source star'' for the event is quite red at baseline but lies well 
below the Large Magellanic Cloud (LMC) giant branch.  On the other hand, the
event itself is quite blue, indicating that the true lensed source is
blue and that the red star at baseline is not being microlensed.
\citet{gbf} proposed that the red star was a foreground M dwarf in the Milky
Way disk and that this M dwarf was in fact the microlens.  While the
optical depth of the Milky Way disk is extremely low, $\tau\sim 10^{-9}$,
the MACHO survey was sufficiently big that one should expect of order
one such event.

{\it Hubble Space Telescope (HST)} observations carried out by 
\citet{alcockprelim} and further analyzed by \citet{alcock01} 
virtually proved this conjecture.  The {\it HST} Planetary Camera
images taken 6.3 years after the event clearly show two stars,
one blue and one red, separated by 134 mas.  The chance that an unrelated
foreground M dwarf would lie so close to a given microlensed source
is only $\sim 10^{-4}$.  Even the conditional probability of such
an alignment, given that a red star was known to lie within the
$\sim 2''$ MACHO PSF, is only a few percent.  \citet{alcock01} therefore
concluded that the red dwarf was in fact the lens.

However, this identification turned out to compound rather than resolve
the puzzle of MACHO-LMC-5.  The lightcurve of this event shows a clear 
asymmetry about its peak (see Fig.~1).  Such asymmetries can be
induced by parallax effects due to the Earth's orbital motion
\citep{gould92}, as was first observed by \citet{al1} and subsequently
in about a dozen other events.  Measuring the parallax effect yields
the (two-dimensional) velocity projected onto the observer plane, 
$\tilde \bv$, of the lens relative to the source.  The lens mass and
lens-source relative parallax are then given by,
\begin{equation}
M = {\tilde v\mu_\rel t_\e^2\over 4 G},
\qquad \pi_\rel = {\mu_\rel\over\tilde v},
\label{eqn:mpi}
\end{equation}
where $t_\e$ is the Einstein crossing time (measured during the event),
and $\mu=134\,{\rm mas}/6.3\,{\rm yr} = 21.4\pm 0.7\,\rm mas\,yr^{-1}$
(measured from the {\it HST} image).  Combining these various 
measurements, \citet{alcock01} found,
\begin{equation}
M = 0.036^{+0.009}_{-0.004}\,M_\odot,\qquad
\pi_\rel = 5\pm 1\,{\rm mas}, \qquad (\rm microlensing),
\label{eqn:mpieval}
\end{equation}
the latter value being equivalent to a lens distance of $D_l\sim 200\,$pc.

Of course, the very fact that the lens could be seen would seem to
argue against such a small, substellar mass.  Indeed, from the observed
color and magnitude of the lens, \citet{alcock01} estimated its mass
and distance to be, 
\begin{equation}
M = 0.11\pm 0.02\,M_\odot,\qquad
D_l = 650\,\pm 190\,{\rm pc}\qquad (\rm photometric).
\label{eqn:mpieval2}
\end{equation}

One way out of this conflict would be simply to assume that the
parallax measurement was wrong.  With a measured timescale of 
$t_\e\sim 30\,$days, MACHO-LMC-5 was about a factor
three shorter than any other event for which there is a reliable
parallax.  Moreover, the parallax detection was only at about the
$5\,\sigma$ level, small enough that it could conceivably be the
product of small and unrecognized systematic errors.  However,
\citet{alcock01} argued that the parallax measurement could not
be so easily dismissed because the direction of $\tilde \bv$
derived from this measurement agreed with the measured direction
of $\bmu_\rel$ to within about $6^\circ$.  The chance of such
an agreement, if the parallax measurement were indeed spurious,
is only about 3\%.  Thus, a decade after its discovery, the
event remains truly a puzzle.

Here I show that the parallax solution for MACHO-LMC-5 is subject to a
four-fold degeneracy.  While two of these solutions are
virtually identical to their counterparts, the other two lead
to quite different estimates of the mass and distance.  One pair of
solutions is equivalent to the solution reported by \citet{alcock01}.
The other pair yields both a larger mass and larger distance. I show that
these are consistent with the photometric determinations.

\section{Microlens Parallax: The Geocentric Worldview
\label{sec:geo}}

While it is customary to fit for microlens parallaxes in the
frame of the Sun, it is actually possible to stay much closer
to the data if one adopts the geocentric point of view, which
is illustrated in Figure~\ref{fig:geo}.  This
can be important, especially in cases like the present one in
which the parallax is only weakly detected.  

Let $\bs(t)$ be the Earth-to-Sun vector in units of AU in the heliocentric
frame.  Let $t_p$ be some fixed time, in practice a time very close
to the time $t_0$ of the peak of the event as seen from the Earth, and 
evaluate the derivative of $\bs(t)$ at this time,
\begin{equation}
\bv_p = {d\bs\over d t}\bigg|_{t_p}.
\label{eqn:bvp}
\end{equation}
Then in the geocentric frame (and relative to its position at $t_p$), 
the Sun has a positional offset (see inset to Fig.~\ref{fig:geo}),
\begin{equation}
\Delta\bs(t) = \bs(t) - (t-t_p)\bv_p - \bs(t_p).
\label{eqn:ageocentric}
\end{equation}
Consider now observations toward an event at some given celestial
coordinates, and define $\bn$ and $\be$ as the unit vectors pointing
north and east.
The projected position of the Sun in the adopted frame will then be
\begin{equation}
(s_n,s_e) = (\Delta\bs\cdot\bn,\Delta\bs\cdot\be).
\label{eqn:anae}
\end{equation}
Note that this coordinate system is right-handed.

Let $(\tau,\beta)$ be the position of the lens relative to the source
in units of the Einstein ring.  Explicitly,
\begin{equation}
\tau(t) = {t-t_0\over t_\e} + \delta\tau,\qquad
\beta(t) = u_0 + \delta\beta,
\label{eqn:tauoft}
\end{equation}
where
\begin{equation}
(\delta\tau,\delta\beta) = \pi_\e\Delta\bs = 
(\bpi_\e\cdot\Delta\bs,\bpi_\e\times\Delta\bs),
%[s_n(t)\pi_{\e,N} + s_e(t)\pi_{\e,E},- s_n(t)\pi_{\e,E} + s_e(t)\pi_{\e,N}]
\label{eqn:betaoft}
\end{equation}
%\begin{equation}
%\tau(t) = {t-t_0\over t_\e} + \delta\tau,\qquad
%\delta\tau = s_n(t)\pi_{\e,N} + s_e(t)\pi_{\e,E}
%\label{eqn:tauoft}
%\end{equation}
%\begin{equation}
%\beta(t) = u_0 + \delta\beta,\qquad
%\delta\beta= - s_n(t)\pi_{\e,E} + s_e(t)\pi_{\e,N},
%\label{eqn:betaoft}
%\end{equation}
$t_\e$ is the Einstein crossing time, and $u_0$ is the
lens-source separation at $t_0$. More explicitly, $(\delta\tau,\delta\beta)=
[s_n(t)\pi_{\e,N} + s_e(t)\pi_{\e,E},- s_n(t)\pi_{\e,E} + s_e(t)\pi_{\e,N}]$.
I define $(\tau,\beta)$ to be
also right-handed, so that if $u_0>0$ then the lens is passing
the source on its right as seen from the Earth.  These equations
serve to define the ``vector microlens parallax'' 
$\bpi_\e = (\pi_{\e,N},\pi_{\e,E})$, whose
magnitude $\pi_\e = |\bpi_\e|$ gives the projected size of the
Einstein ring, $\tilde r_\e = {\rm AU}/\pi_\e$, and whose direction
gives the direction of the lens relative to the source as seen
in the adopted frame.   That is, at $t=t_p$, 
$d(\tau,\beta)/dt=(1,0)/t_\e$.  So if the lens is going due north
$[\bpi_\e = (\pi_{\e,N},0)]$,
the parallax deviation $(\delta\tau,\delta\beta)=(s_n,s_e)\pi_\e$,
while if it is going due east,
 $(\delta\tau,\delta\beta)=(s_e,-s_n)\pi_\e$, which, since
both $(\tau,\beta)$ and $(s_n,s_e)$ are right-handed, 
are the proper behaviors. 

There are several advantages to using these variables.  Most significantly,
when the event is fit including the parallax effect, the parameters
$t_0$, $u_0$, and $t_\e$, will come out to be very similar to their
values when it is fit without parallax.  That is, these parameters
are given directly by the data and do not depend on the parallax
model.  This can be very important for cases in which the parallax
is not strongly constrained.  In such cases, the trajectory relative
to the Sun will also not be well constrained, so the errors in 
$t_0$, $u_0$, and $t_\e$ in the heliocentric frame will be huge.  But these
errors will also be extremely correlated, since whatever values one
adopts, they must conspire to produce exactly the right 
peak amplitude at exactly the right time, and passing at exactly the right rate
as seen from the Earth.  The downside is that at the end of the day,
one must still convert to heliocentric coordinates in order to extract some
of the parameters.  However, it is actually better to
perform this step separately so as to be able to understand the
various sources of uncertainty in the final measurement.

Also, note that I am fitting for $\bpi_\e$ rather than
$\tilde r_\e\equiv \au/\pi_\e$ or 
$\tilde \bv\equiv(\tilde r_\e/\tilde t_\e)(\bpi_\e/\pi_\e)$.
As with trigonometric parallaxes, microlens parallaxes are much better
behaved than their inverse quantities, particularly when they are near
zero.  See also \citet{natural}.  

\section{Parallax Fits
\label{sec:parfit}}

\begin{figure}
\plotone{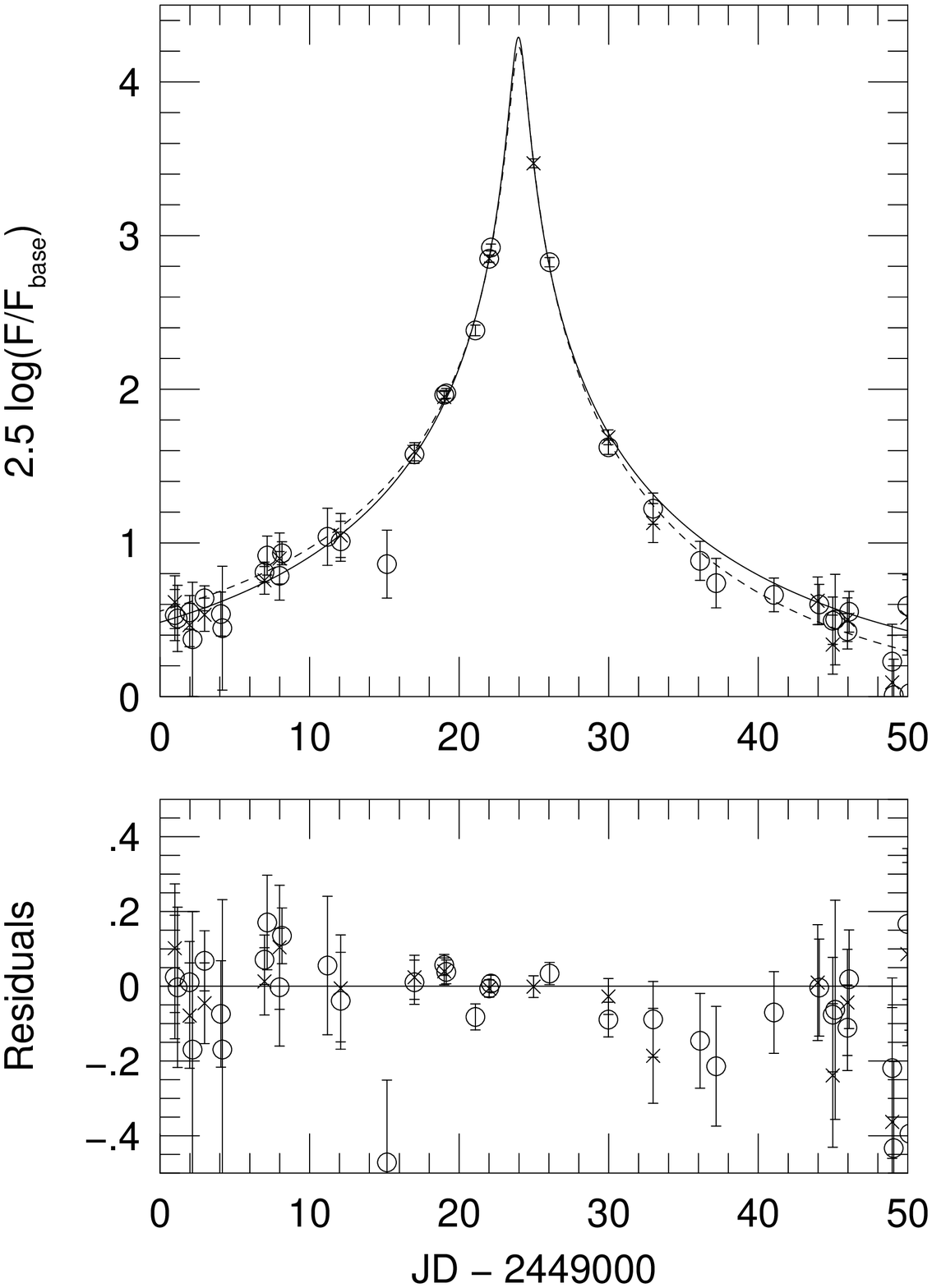}
\caption{Lightcurve of MACHO-LMC-5 for MACHO red ({\it circles}) and
blue ({\it crosses}) filters.  Fluxes are normalized to 
$F_{\rm base}\equiv F_s+F_b$ for red and are aligned to the same system
via a linear transformation for blue.  The solid curve is the best
fit without parallax and shows clear asymmetric residuals, which are
characteristic of parallax.  See lower panel.  The dashed curve shows
the best fit with parallax.
\label{fig:lc}}
\end{figure}

\begin{figure}
\plotone{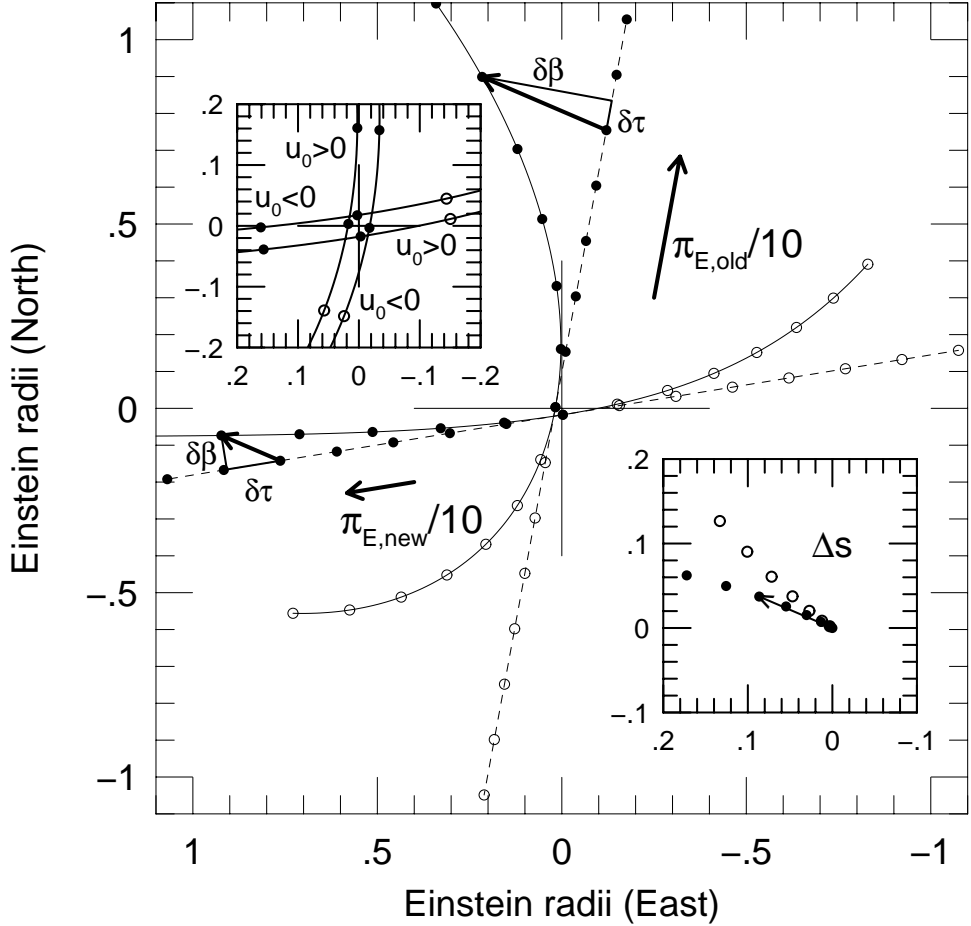}
\caption{The geocentric view of microlens parallax for the old and new
solutions of the event MACHO-LMC5, discussed in this paper.  The inset 
at the lower right shows
$\Delta \bs(t)$, the apparent path of the Sun (in AU), relative
to its position in the geocentric frame defined by the Earth's motion
at the peak of the event, and projected onto the plane of the
sky in (east, north) coordinates.  {\it Open circles} are for 
$t<t_0$ and {\it filled circles} are for $t\geq t_0$.  Point separation
is 5 days. The {\it dashed lines} represent the
path of the lens relative to the source ({\it central cross}) in Einstein
radii and in absence of parallax, with $(\tau,\beta)=([t-t_0]/t_\e,u_0)$.
In both cases the lens passes the source on its right, so $u_0>0$.
That is, the $(\tau,\beta)$ coordinate system is right-handed.
The effect of parallax is to displace the lens by 
$(\delta\tau,\delta\beta) = \pi_\e\Delta \bs=
(\bpi_\e\cdot \Delta\bs,\bpi_\e\times \Delta\bs)$
to $(\tau,\beta)=([t-t_0]/t_\e+\delta\tau,u_0+\delta\beta)$
({\it solid curves}).  The
displacement vector is shown explicitly for
$t=t_0+25\,{\rm days}$.  Note that for both solutions, the offsets are
parallel to $\Delta\bs$ and their magnitudes are proportional to $\pi_e$.
However, their decomposition into $\delta\tau$ and $\delta\beta$ is
very different because the $\tau$ direction (the direction of motion and
so also the direction of $\bpi_\e$) is different.  The inset at the upper
left displays all four degenerate solutions, including the two positive
$u_0$ solutions shown in the main figure as well as the two negative
$u_0$ solutions.  See Table 1.
\label{fig:geo}}
\end{figure}

I begin with the data set obtained from the MACHO web site
(http://www.macho.mcmaster.ca).  The event has a total of 265 points
in $R_M$ and 352 in $B_M$.  I eliminate 3 outliers (all at baseline)
and find that the remaining points have a
$\chi^2/{\rm dof}=0.65$ in each filter separately.  This indicates
that the errors have been overestimated.  In principle, one might
under these circumstances renormalize the errors by a factor
$\sqrt{0.65}=0.81$.  However, the great majority of the points
are at baseline where the event is extremely faint, whereas most
of the information of immediate interest comes from the highly
magnified portions of the event, where the error corrections are not
likely to be the same as for the baseline
points.  Hence I do not renormalize.
As mentioned in the Introduction, the fit to a standard
\citet{pac86} curve, with 7 parameters, $t_0$, $u_0$, $t_\e$,
$f_{s,R}$, $f_{b,R}$, $f_{s,B}$, and $f_{b,B}$,
\begin{equation}
f_i(t) = f_{s,i}A(u[t]) + f_{b,i},\qquad A(u) = {u^2+2\over u\sqrt{u^2+4}},
\label{eqn:foft}
\end{equation}
where $[u(t)]^2 = u_0^2 + (t-t_0)^2/t_\e^2$,  
shows clear asymmetric residuals.  See Figure~\ref{fig:lc}.

\begin{figure}
\plotone{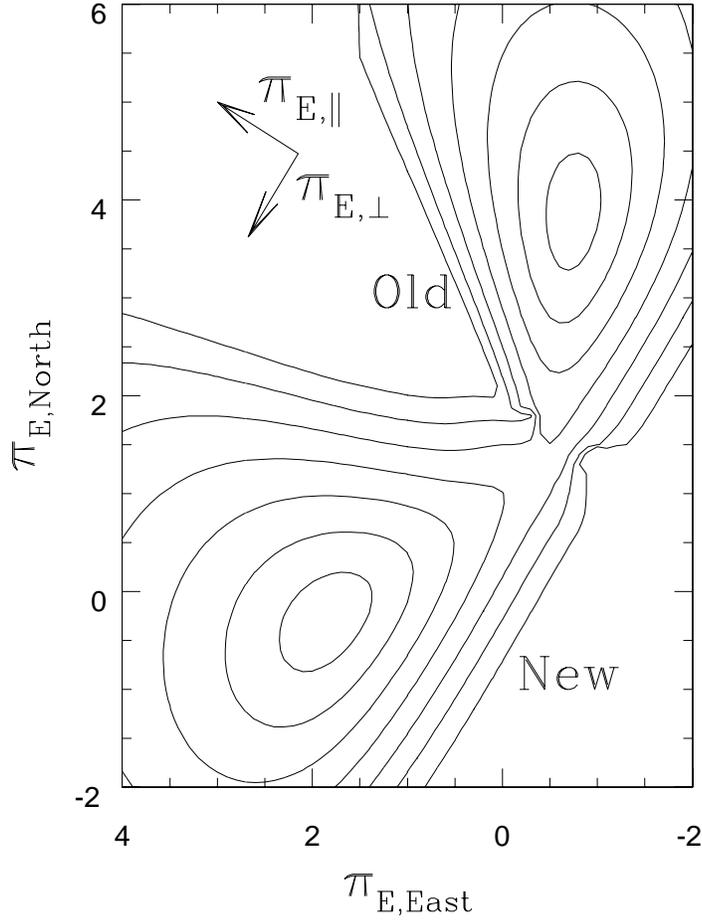}
\caption{Likelihood contours in the $\bpi_\e$ plane shown at
$\Delta\chi^2=1$, 4, 9, 16, 25, 36, and 49 relative to the minimum.
There are two solutions, one to the northwest previously found by
\citet{alcock01}, and a new one to the southeast found in this paper.
The directions of positive $\pi_{\e,\parallel}$ and $\pi_{\e,\perp}$
are shown as a ``corner''.  The offset between the two solutions is
almost exactly aligned with the $\pi_{\e,\perp}$ 
direction, which is perpendicular to the
Earth's acceleration vector at the event maximum.  The diagram shows
only the $u_0<0$ solutions, but the contours for $u_0>0$ are virtually
identical.  Note that $\bpi_\e$ is dimensionless, since 
$\pi_\e = \au/\tilde r_\e$.
\label{fig:piecontours}}
\end{figure}

I then add two additional parameters, $\pi_{\e,N}$ and $\pi_{\e,E}$,
which enter $[u(t)]^2=[\tau(t)]^2 + [\beta(t)]^2$ through equations
(\ref{eqn:tauoft}) and (\ref{eqn:betaoft}).  At first, I use 
the no-parallax solution as my seed.  The code converges to a solution
that is inconsistent with the results of \citet{alcock01}.  See
Table 1.  I therefore explore a densely sampled grid over the rectangle
$-2\leq\pi_{\e,N}\leq 6$, $-2\leq\pi_{\e,E}\leq 4$.  The likelihood
contours of this search are shown in Figure~\ref{fig:piecontours}.

Figure~\ref{fig:piecontours} 
has a number of notable features.  First, of course, it has
two solutions.  The second solution (to the northwest) is the same
as the one found by \citet{alcock01}.  As shown in Table 1, the
two solutions differ in $\chi^2$ by less than 0.1.  Hence, they are
truly degenerate.  Second, the high $\chi^2$ contours to the 
southwest of the two solutions tend toward continuous straight lines
with a position angle of about $149^\circ$ (North through East).
This is almost exactly perpendicular to the acceleration of the
Earth (projected onto the plane of the sky, which has an amplitude
of $0.52\,\kms\,\rm day^{-1}$ and a position angle of 
$238.\hskip-2pt^\circ 3$.
That is, these contours derive from the parallax asymmetry that is
due to the acceleration of the Earth along the direction of lens motion
and which is clearly visible in Figure~\ref{fig:lc}.
For events with weak parallax, one obtains only this one-dimensional
information about the parallax \citep*{gmb}.  Evidently, MACHO-LMC-5
is relatively close to this situation (as one would expect from its
brevity), but Figure~\ref{fig:piecontours} 
shows that this event lies in a region
of the $\bpi_\e$ diagram that is {\it beyond} this continuous
degeneracy.  

\section{Four-fold Degeneracy
\label{sec:fourfold}}

In fact, there are not just two solutions, but four.  The
other two solutions are obtained from the first two by
first substituting $u_0\rightarrow -u_0$ and then making
very slight adjustments to the other parameters.  This
degeneracy was discovered by \citet*{smp}.  See Table 1.

\subsection{Analytic Description
\label{sec:analytic}}

To understand the nature of this degeneracy, I extend the approach
of \citet{smp} by Taylor expanding $\bu$, the vector position of the
lens relative to the source in the Einstein ring,
\begin{equation}
\bu = \bu_0 + \bomega t + \pi_\e \biggl(
{1\over 2} \balpha t^2 + {1\over 6}\bj t^3 + \ldots \biggr)
\label{eqn:vecu}
\end{equation}
where $\bu_0$ is the vector impact parameter, $\bomega$ is the
vector inverse timescale (i.e., $\omega=t_\e^{-1}$ with direction
given by the lens-source relative motion), 
and $\balpha$ and $\bj$ are the apparent
acceleration and jerk of the Sun relative to the Earth, both
divided by an AU.  Note
that $\bomega$, $\balpha$, and $\bj$ are all evaluated at $t=0$
and that all are two-dimensional vectors.
I impose $\bu_0\cdot \bomega=0$, which is equivalent to assuming
that $t_0$ (the time of closest approach) can be directly determined
from the lightcurve and so does not require an additional parameter.
Squaring equation~(\ref{eqn:vecu}) yields,
\begin{equation}
u^2 = \sum_{i=0}^\infty C_i t^i,
\label{eqn:usquared}
\end{equation}
where,
\begin{equation}
C_0 = u_0^2,\quad C_1=0,\quad C_2= -\alpha u_0\pi_{\e,\perp} + t_\e^{-2}
\label{eqn:C02}
\end{equation}
\begin{equation}
C_3 =  \alpha {\pi_{\e,\parallel}\over t_\e} + 
{1\over 4}\alpha^2 t_\e u_0 \bpi_\e\times\bpi_j,
\label{eqn:C3}
\end{equation}
\begin{equation}
C_4 = {\alpha^2\over 4}(\pi_\e^2 + \bpi_j\cdot\bpi_\e) 
+ {1\over 12}{\Omega_\oplus^2\over\alpha}u_0 \pi_{\e,\perp},
\label{eqn:C4}
\end{equation}
where I have introduced the ``jerk parallax'',
\begin{equation}
\bpi_j \equiv {4\over 3}{\bj\over \alpha^2 t_\e},
\label{eqn:jerkpar}
\end{equation}
and where the subscripts ``$\parallel$'' and ``$\perp$'' indicate
components parallel and perpendicular to the acceleration $\balpha$.
Note that I have made use of the fact that the Earth's orbit is basically
circular to approximate the derivative of the jerk as 
$-\Omega_\oplus^2\balpha$, where $\Omega_\oplus=2\pi\,\rm yr^{-1}$.

If one has found one set of parameters $(u_0,t_\e,\bpi_\e)$ that
fit the light curve, then one can empirically determine the
constants $(C_0, C_2, C_3, C_4)$.  Any other parameter combination
$(u_0',t_\e',\bpi_\e')$ that reproduces these constants will then
provide an equally good fit to the lightcurve (at least to fourth
order in $t$).  In principle, one could solve 
equations~(\ref{eqn:C02})-(\ref{eqn:C4}) numerically, but more
physical insight can be gained by solving them algebraically in
two relevant limits.

\subsection{The Constant-Acceleration Degeneracy
\label{sec:smp}}

If we ignore the jerk and jerk-derivative terms, then 
equations~(\ref{eqn:C3}) and (\ref{eqn:C4}) become,
\begin{equation}
C_3 =  \alpha {\pi_{\e,\parallel}\over t_\e},\quad
C_4 = {\alpha^2\over 4}\pi_\e^2.
\label{eqn:C34}
\end{equation}
\citet{smp} showed that equations~(\ref{eqn:C02}) and (\ref{eqn:C34})
have two degenerate solutions, which are roots of the cubic
equation,
\begin{equation}
x^3 -2 C_2 x^2 + [C_2^2-(\alpha u_0\pi_\e)^2]x + C_3^2 u_0^2 = 0,
\label{eqn:cubic}
\end{equation}
where $x=t_\e^{-2}$, and where all the parameters are evaluated
at one solution.
While the roots of cubics
are normally ungainly, in this case one of the roots is known
from having found the solution to the event numerically.  I label
this solution with unprimed variables $(t_\e,\bpi_\e,u_0)$ and
label the second (so far unknown) solution with primed variables
$(t_\e',\bpi_\e',u_0')$.
Hence, by dividing the cubic by the known root, it can be reduced
to a quadratic, which is easily solved.  I find,
\begin{equation}
t_\e' = \Biggl[ (1-2\epsilon_\perp)
{1 + \sqrt{1 + [2\epsilon_\parallel/(1-2\epsilon_\perp)]^2}\over 2}
\Biggr]^{-1/2} t_\e \sim (1+\epsilon_\perp)t_\e,
\label{eqn:teprime}
\end{equation}
where $\bepsilon \equiv \alpha t_\e^2 u_0\bpi_\e$. 
As shown by \citet{smp},
the other second-model parameters follow easily from this
evaluation,
\begin{equation}
u_0'=-u_0,\quad
\pi_\e' = \pi_\e,\quad
\pi_{\e,\parallel}' = {t_\e'\over t_\e}\pi_{\e,\parallel}.
\label{eqn:othereval}
\end{equation}

Note that the relations between the timescales for the two pairs
of solutions shown in Table 1 are
almost exactly as predicted by equation~(\ref{eqn:teprime}): from
the $u_0<0$ solutions, one would predict the timescales of the
$u_0>0$ solutions to be 32.2137 and 32.7270 days respectively.
The actual timescales differ fractionally from these values by 
$<10^{-4}$.  

\subsection{The New Jerk-Parallax Degeneracy
\label{sec:jerk}}

To elucidate the new degeneracy, I consider the limit $u_0\rightarrow 0$.
Equations~(\ref{eqn:C3}) and (\ref{eqn:C4}) then become,
\begin{equation}
C_3=\alpha{\pi_{\e,\parallel}\over t_\e},\qquad
C_4 = {\alpha^2\over 4}(\pi_\e^2 + \bpi_\e\cdot\bpi_j),
\label{eqn:C34p}
\end{equation}
for which the solution is,
\begin{equation}
\pi_{\e,\parallel}' = \pi_{\e,\parallel},\qquad
\pi_{\e,\perp}' = -(\pi_{\e,\perp}+\pi_{j,\perp}),
\label{eqn:pisols}
\end{equation}
and $t_\e' = t_\e$.

Note that for the special case $\pi_{\e,\perp} = -\pi_{j,\perp}/2$, 
equation~(\ref{eqn:pisols}) implies that $\bpi_\e'=\bpi_\e$.  That
is, there is no degeneracy.  

\subsection{Full Solution
\label{sec:full}}

 It is now possible to solve for the adjustments to the other
parameters to first order in $u_0$.  For example, from the full
expression for $C_2$ in equation~(\ref{eqn:C02}), one finds,
\begin{equation}
\Delta t_\e \simeq -{\alpha t_\e^2\over 2}
(\pi_{\e,\perp}\Delta u_0 + u_0\Delta\pi_{\e,\perp})t_\e.
\label{eqn:Deltate}
\end{equation}
Note that for the degeneracy originally identified by \citet{smp},
for which $\Delta u_0 = -2u_0$ and $\Delta \pi_{\e,\perp}=0$,
this equation reduces to the first-order evaluation of 
equation~(\ref{eqn:teprime}).  However, it also correctly
predicts the timescale differences of all four degenerate
solutions listed in Table 1.  For example,  the difference
between the $t_\e$'s of the two $u_0>0$ (or the two $u_0<0$) 
solutions is predicted to
be $\Delta t_\e= 0.0134 t_\e$, whereas the actual differences are 
$0.0114 t_\e$ and $0.0158t_\e$, respectively.

\subsection{Comparison with Zero-Parallax Solution}

The fifth line of Table 1 shows the best fit with the
parallax enforced to be zero.  Most of the seven parameters
differ from their counterparts in the four parallax solutions
by amounts that are large compared to the internal scatter of
these solutions.  This seems to contradict the claim made
in \S~\ref{sec:geo} that the advantage of the ``geocentric
worldview'' is that the fit parameters do not substantially
change with and without parallax.  In fact, for high-magnification
events with faint sources (like MACHO-LMC-5), the parameters that
are well constrained by the fits are not $u_0$, $t_\e$, and $F_s$,
separately, but rather the parameter combinations
$F_{\rm max}=F_s/u_0$, $t_{\rm eff}=u_0 t_\e$, and their product
$F_{\rm max}t_{\rm eff} = F_s t_\e$.  To test this, I show in the
sixth line of Table 1, the best fit zero-parallax solution with 
$F_{s,R}$ constrained to the value of the first parallax solution.
The parameters of this solution show much better agreement with
those of the parallax solutions, while $\chi^2$ is increased by
less than unity (for one more degree of freedom).  Hence,
the geocentric worldview is confirmed.

%\begin{equation}
%\label{eqn:}
%\end{equation}

\section{Lens Mass and Distance
\label{sec:massdist}}

\begin{figure}
\plotone{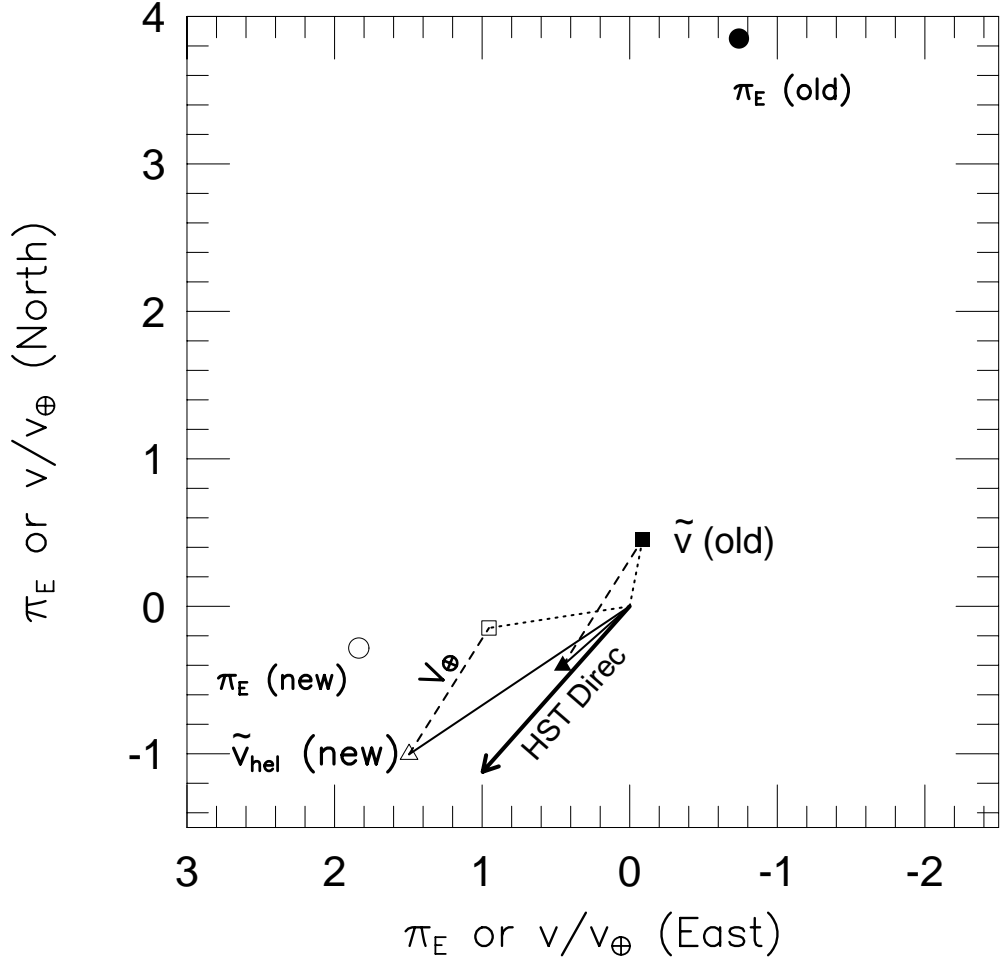}
\caption{
Transformation from directly observable to physically relevant quantities.
Filled and open symbols are for the old 
\citep{alcock01} and new (present paper) solutions, respectively.
Geocentric vector parallaxes $\bpi_\e$ ({\it circles}) are measured from the
lightcurve together with standard microlensing parameters 
(see Fig.~\ref{fig:piecontours}).  Geocentric projected velocities 
$\tilde \bv$ ({\it squares}) are found by inverting the magnitudes of
$\bpi_\e$, but keeping the same directions.  See eq.~(\ref{eqn:geovel}).
The Earth velocity at event maximum ({\it dashed lines}) is added to each of
these to obtain the heliocentric projected velocities $\tilde \bv_{\rm hel}$.
See eq.~(\ref{eqn:helvel}).  The direction of these ({\it thin solid lines})
should be the same as that of the lens-source relative proper motion
$\bmu_\rel$ measured by {\it HST}, which is shown as a bold line.
Note that $\bpi_\e$ is dimensionless, since $\pi_\e = \au/\tilde r_\e$.
\label{fig:projvel}}
\end{figure}

To compare the direction of motion of the parallax solution
with that of the {\it HST} proper-motion measurement,
$\theta_{HST}=138.\hskip-2pt ^\circ 4$
(which reflects the time-averaged motion of the Earth and so is
basically heliocentric), the parallax measurement must be converted
from geocentric to heliocentric coordinates.  This requires two
steps.  First, the parallax must be converted to the geocentric
projected velocity,
\begin{equation}
\tilde \bv = {{\rm AU}\over t_\e}\,{\bpi_\e\over \pi_\e^2}.
\label{eqn:geovel}
\end{equation}
Next, one must add the Earth's instantaneous velocity at $t_0$
to obtain the heliocentric projected velocity,
\begin{equation}
\tilde \bv_{\rm hel} = \tilde \bv + \bv_\oplus(t_0).
\label{eqn:helvel}
\end{equation}
Finally, since the projected Einstein radius, $\tilde r_\e=\tilde v t_\e$,
is the same in both frames, the heliocentric Einstein crossing time
is,
\begin{equation}
t_{\e,\rm hel} = {\tilde v\over \tilde v_{\rm hel}}t_\e.
\label{eqn:tehel}
\end{equation}
These steps are illustrated in Figure~\ref{fig:projvel}.   Also
shown in this figure is the direction $\bmu_\rel$ as measured by
{\it HST}.  Note that both parallax solutions are
reasonably aligned with this direction, although the new solution
is about twice as far from it as the original \citet{alcock01} solution.
See Table 1.

\begin{figure}
\plotone{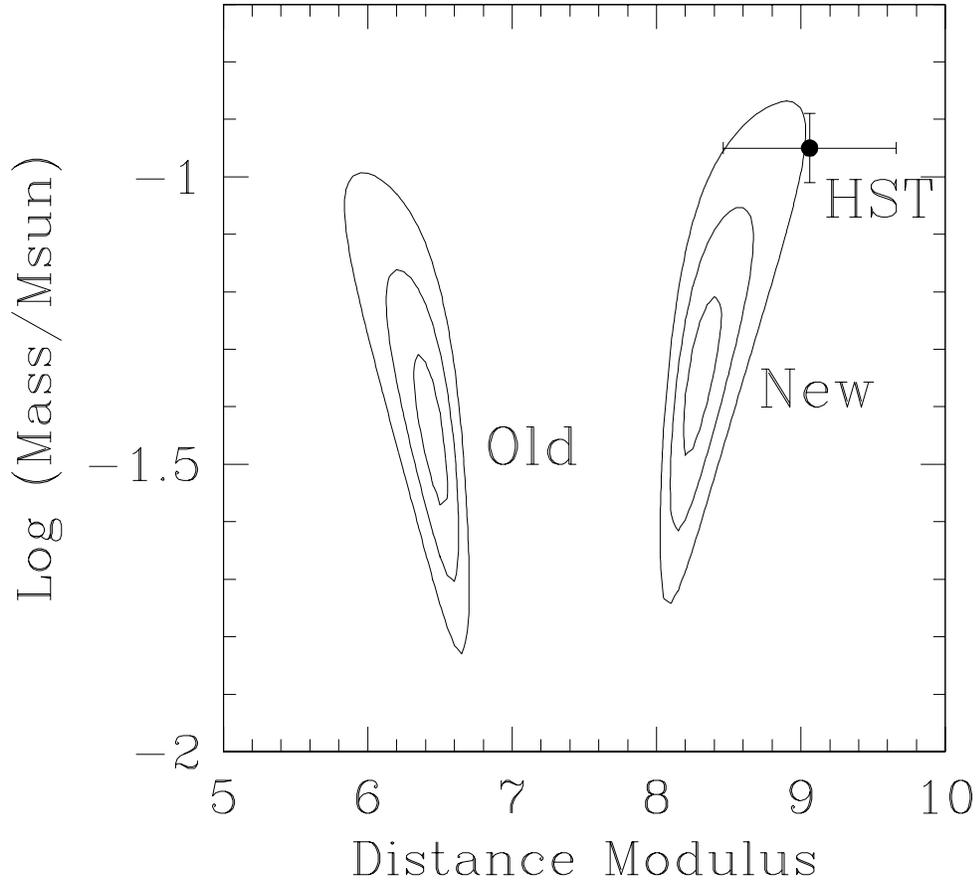}
\caption{Likelihood contours ($\Delta\chi^2=1$, 4, and 9)
for lenses of mass $M$ and distance modulus $5\log [D_l/10\,\rm pc])$,
where $D_l$ is the lens distance.  
The new solution lies to the
right.  The photometrically determined mass and distance
({\it filled circle}) are reasonably consistent with the new solution but
are inconsistent with the old one.
\label{fig:md}}
\end{figure}

I now combine the {\it HST} proper motion with each of the two
parallax solutions to produce likelihood contours in lens mass and
distance.  For simplicity, and because the {\it HST} angular measurement
error ($1.\hskip-2pt ^\circ 0$)
is so much smaller than the difference between it and either
parallax solution, I treat the {\it HST} $\bmu_\rel$ as a constraint,
i.e., as having zero error.  For each lens trial distance 
$D_l\simeq {\rm AU}\pi_\rel^{-1}$,  I derive a heliocentric projected
velocity, 
$\tilde \bv_{\rm hel} = {\rm AU}\bmu_\rel \pi_\rel \simeq D_l\bmu_\rel$.
I convert this to a geocentric projected velocity $\tilde \bv$ by inverting
equation~(\ref{eqn:helvel}).
Next, for each lens mass $M$, I infer a microlens parallax,
\begin{equation}
\pi_\e = \sqrt{\pi_\rel\over\kappa M},\qquad \kappa = 
{4 G\over c^2\,\rm AU}\sim 8.1{{\rm mas}\over M_\odot},
\label{eqn:pieeval}
\end{equation}
and determine the geocentric Einstein timescale by 
$t_\e = {\rm AU}(\pi_\e \tilde v)^{-1}$.  Finally, I find the
vector geocentric microlens parallax $\bpi_\e$ by inverting 
equation~(\ref{eqn:geovel}).  For each $(M,D_l)$ pair, I then
minimize $\chi^2$ holding the parameters $(\bpi_\e,t_\e)$ fixed
at the values thus calculated.  Figure~\ref{fig:md} shows the
resulting likelihood contours for the two sets of solutions.  Each
set of contours is shown relative to its own local minimum,
which are offset from the global minimum (without the proper-motion
constraint) by $\Delta\chi^2=2.0$
and $\Delta\chi^2=3.8$ for the original and new solutions respectively.
Note that while both sets of contours extend up to the hydrogen-burning
limit at the few $\sigma$ level, the old solution does so by moving
toward smaller distances, while the new solution does so by moving
toward larger distances.  The best estimate for the lens mass and distance
derived by \citet{alcock01} from its {\it HST} color and magnitude is
shown by a circle with error bars.  This is reasonably consistent 
(at the $2.5\,\sigma$ level) with
the new solution but not the old one.

The reason for the divergent behaviors toward higher mass 
of the two sets of contours can be understood from Figure~\ref{fig:projvel}.
As the distance is increased, so is the heliocentric projected speed,
$\tilde v_{\rm hel}\simeq D_l\mu_\rel$.  From Figure~\ref{fig:projvel},
this increases the magnitude of $\tilde v$ for the new solution, but
decreases it for the old solution.  Since the geocentric timescale $t_\e$
is basically fixed directly by the lightcurve and so is approximately
the same for all solutions, the mass 
$M\propto \tilde v_{\rm hel}\mu_\rel t_\e^2$
moves in tandem with the heliocentric projected speed.  Hence, the
mass is correlated with the distance for the new solution but
anticorrelated for the old solution.

\section{Discussion
\label{sec:discuss}}

\subsection{Levels of Parallax Degeneracy
\label{sec:levels}}

 The Taylor expansion of the (squared) lens-source separation,
equations~(\ref{eqn:usquared})-(\ref{eqn:C4}), allows one to
understand analytically the various levels of microlensing parallax
degeneracy.
If there is only enough information in the lightcurve to measure the
first two terms (out to $C_2$), then there is no information at all about
$\pi_\parallel$, while $\pi_\perp$ is completely degenerate with the
timescale $t_\e$.  Hence, nothing can be learned about the parallax,
and it is usually assumed to vanish.

If the next ($C_3$) term can be measured (and if the second term in
eq.~[\ref{eqn:C3}] can be ignored) then this equation allows a
determination of $\pi_\parallel$, but $\pi_\perp$ remains indeterminate.
This is the degeneracy identified by \citet{gmb}.  Taking account
of the second term in equation~(\ref{eqn:C3}) leaves a linear
continuous (i.e., line-like) degeneracy in the $\bpi_\e$ plane, but
rotates this line by an angle,
\begin{equation}
\theta_\pi = \tan^{-1}{\eta\pi_{j,\parallel}\over 1+\eta\pi_{j,\perp}},
\qquad\eta\equiv {1\over 4}{u_0\alpha t_\e^2}.  
\label{eqn:thetapi}
\end{equation}
Since typically $\eta\pi_j\ll 1$,
this angle is also usually very small, but it can be significant in
some cases (see \S~\ref{sec:implications}).

If $C_4$ can be measured, then the continuous degeneracy is broken,
but it is replaced by a four-fold discrete degeneracy.  In this
case, all four solutions lie along the line of the continuous
degeneracy described in the previous paragraph.  The transition
from the continuous to the discrete degeneracy is itself continuous.
If $C_4$ is measured, but with only modest significance, then the
error ellipses for each of the four solution will be extremely
elongated along the line of continuous degeneracy.  In this case,
the two solutions with $u_0>0$ may merge.  Similarly for the two
solutions with $u_0<0$.  However, the $u_0>0$ and $u_0<0$ solutions cannot
merge with each other unless $u_0^2$ is consistent with zero. 
To break the discrete degeneracy requires sensitivity to the higher order
terms, which is typically obtained only in relatively long events.

\subsection{Broad Implications
\label{sec:implications}}

For which microlensing events is the jerk-parallax degeneracy 
likely to be important?  To address this question, 
it is best to think in terms of
the ``inverse projected velocity'', $\bLambda$,
\begin{equation}
\bLambda \equiv {\tilde \bv\over \tilde v^2} ={t_\e\over {\rm AU}}\bpi_\e.
\label{eqn:bLambdadef}
\end{equation}
(I choose ``$\bLambda$'' for this quantity because it looks like an
inverted ``$\bv$''.)
Since $t_\e$ is virtually the same for all solutions, $\bLambda$ is
basically just a linear rescaling of $\bpi_\e$.  Its usefulness derives from
the fact that the projected velocity (and so its inverse) is related
solely to the kinematics of the event and thus is independent of the
mass.  Expressed in terms of this quantity, equation~(\ref{eqn:pisols})
becomes
\begin{equation}
\Lambda_\parallel'=\Lambda_\parallel,
\qquad
\Lambda_\perp'=-(\Lambda_\perp + \tilde v_j^{-1}),
\label{eqn:bLambdasols}
\end{equation}
where,
\begin{equation}
\tilde v_j \equiv {3\over 4}\,{\alpha^3\over\balpha\times\bj}{\rm AU}.
\label{eqn:bLambdajdef}
\end{equation}
(Note that since $\balpha$ and $\bj$ are 2-dimensional vectors,
the denominator is a signed scalar.  See also eq.~[\ref{eqn:C3}].)\ \
If we approximate the Earth's orbit as circular, then $\tilde v_j$
may be evaluated,
\begin{equation}
\tilde v_j  = -{3\over 4}\,
{(\cos^2\psi \sin^2\beta_\ec + \sin^2\psi)^{3/2}\over \sin\beta_\ec}\,v_\oplus,
\label{eqn:Lambdajeval}
\end{equation}
where $\beta_\ec$ is the ecliptic latitude of the event, $\psi$ is the
phase of the Earth's orbit at $t_0$ relative to opposition, and
$v_\oplus=30\,\kms$ is the speed of the Earth.

Note first that for LMC events, $\tilde v_j \simeq (3/4)v_\oplus$.
For the great majority of LMC events, those with lenses in the
Galactic halo or the LMC itself, $\tilde v\gg v_\oplus$.  Hence, 
$\tilde v_j^{-1}\gg \Lambda$, 
so it would seem at first sight that the degeneracy
would be very important.  In fact, for events with $\tilde v\gg v_\oplus$,
parallax effects will not generally be detectable unless the event is quite
long, $\Omega_\oplus t_\e>1$.  Such long events would most likely not 
in fact be degenerate because the corresponding high-parallax solution 
would have strong signatures at Taylor-expansion orders beyond $t^4$.
It is only the disk-lens events, such as MACHO-LMC-5, for which
this degeneracy is likely to cause confusion.

For events seen toward the bulge, the situation depends very strongly
on the time of year.  For definiteness, consider events toward
Baade's Window, for which $\sin\beta_\ec\sim -0.1$.  At opposition (roughly
the summer solstice), $\tilde v_j = 0.0075\,v_\oplus\sim 0.2\,\kms$, 
which is much
slower than the $\tilde v$ of any plausible event.  
In this case, the reasoning just
given for the majority of LMC events applies, so the degeneracy is
not likely to be important.

At quadrature (roughly the equinoxes) 
$\tilde v_j = 7.5\,v_\oplus\sim 225\,\kms$, 
which is comparable to typical $\tilde v$ for lensing of bulge sources
by disk lenses and is smaller by a factor of a few than typical $\tilde v$
for bulge-bulge lensing.  Hence, this degeneracy could have significant
impact on the interpretation of events peaking near the equinoxes.

At intermediate times, $v_j$ scales roughly as 
$v_j\sim 7.5\,v_\oplus\sin^3\psi$, so that, for example, at 
$\psi=\pm 60^\circ$, $v_j\sim 145\,\kms$, still large enough to cause
significant confusion.

%\begin{equation}
%\label{eqn:}
%\end{equation}

%$\theta=123.\hskip-2pt^\circ 9$, which is not quite in as good agreement

\acknowledgments
I thank the referee, Shude Mao, for suggesting the inclusion of 
Figure 2, which significantly clarifies the adopted geometry.
This work was supported by grant AST 02-01266 from the NSF.

\clearpage

\clearpage


\end{document}